\documentclass[aps,pre,preprint,groupedaddress,showpacs]{revtex4}

\begin{document}


\title{Fluctuation-Dissipation Theorem for the Microcanonical Ensemble}


\author{Marcus V. S. Bonan\c{c}a}
\email[]{marcus.bonanca@physik.uni-regensburg.de}
\affiliation{Institut f\"ur Theoretische Physik, Universit\"at Regensburg, D-93040 Regensburg, Germany}


\date{\today}

\begin{abstract}
A derivation of the Fluctuation-Dissipation Theorem for the microcanonical ensemble is presented using linear response theory. The theorem is stated as a relation between the frequency spectra of the symmetric correlation and response functions. When the system is not in the thermodynamic limit, this result can be viewed as an extension of the fluctuation-dissipation relations to a situation where dynamical fluctuations determine the response. Therefore, the relation presented here between equilibrium fluctuations and response can have a very different physical nature from the usual one in the canonical ensemble. These considerations imply that the Fluctuation-Dissipation Theorem is not restricted to the context of the canonical ensemble, where it is usually derived. Dispersion relations and sum rules are also obtained and discussed in the present case. Although analogous to the Kramers-Kronig relations, they are not related to the frequency spectrum but to the energy dependence of the response function.
\end{abstract}

\pacs{05.20.-y, 05.30.-d, 05.40.-a, 05.70.Ln}
\keywords{fluctuation-dissipation theorem, microcanonical ensemble, linear response theory}

\maketitle

\section{Introduction}

The relation between the fluctuations occurring in a system at equilibrium and dissipation effects dates back to Einstein \cite{einstein} and his theory on Brownian motion. After that, Nyquist \cite{nyquist} derived a relation between the electrical resistance and voltage fluctuations in linear electrical systems. It was realized then by Callen and Welton \cite{callen} that such a relation could be proven for general linear dissipative systems using quantum mechanics. At that moment, the intuition of the authors, as described in the last paragraph of their Introduction, was that the relationship between equilibrium fluctuations and irreversibility would provide a method for a general approach to a theory of irreversibility and, indeed, this was the way pursued by Kubo \cite{kubo1} to achieve the theory of linear response. It is well established now that linear response theory gives a general proof of the Fluctuation-Dissipation Theorem (FDT) which states that the linear response of a given system to an external perturbation is expressed in terms of the fluctuation properties of the system in thermal equilibrium.

Because of this deep relation between the FDT and linear response theory, it is worth noting that the response, as formulated by that theory, is given for any equilibrium ensemble. In other words, the response function can, in principle, be known not only when the system is initially in thermal equilibrium but also in another equilibrium state such as, for example, the microcanonical one. Therefore, the theory is quite general in the sense that the linear response of a system and its equilibrium fluctuations could be related to each other for any kind of equilibrium conditions. Indeed, fluctuation-response relations have been derived even in the context of stochastic systems \cite{deker,hanggi} and non-Hamiltonian deterministic systems \cite{marconi} using linear theory. Perhaps the very first work concerning different equilibrium conditions from the thermal one in Hamiltonian systems is Ref.\cite{bishop}, where the author shows that Kubo's formula can also be derived in the classical microcanonical ensemble as long as the thermodynamic limit is considered. However, for many and different reasons, much more attention was given for the statistical mechanics in the canonical ensemble than in the microcanonical one and the generality of linear response theory concerning different equilibrium conditions was not much explored. Of course, one could argue that the equivalence of the ensembles in the thermodynamic limit would be the reason for focusing just in the canonical ensemble, but recent developments have shown that there are indeed strong motivations to consider different equilibrium situations. For example, a path integral representation for the quantum microcanonical ensemble \cite{lawson} presented a few years ago was motivated by situations where the microcanonical approach may be more appropriate as for the description of systems at low temperatures or with a finite number of particles. The microcanonical ensemble has also been considered in relations between fluctuation and dissipation in systems far from equilibrium like the Crooks relation, where its microcanonical version helps to understand the connection between various of those fluctuation theorems \cite{broeck}. In Ref.\cite{talkner}, a derivation of a microcanonical quantum fluctuation theorem was presented. Considering the work performed by a classical force on a quantum system when it is initially prepared in the microcanonical state, the authors provide a relation that could be accessible experimentally to measure entropies. In the context of nanosystems, where the number of degrees of freedom constituting the environment is not always large enough, the microcanonical ensemble has also been considered. In Ref.\cite{esposito}, a quantum master equation was derived describing the dynamics of a subsystem weakly coupled to an environment of finite heat capacity and initially described by a microcanonical distribution. Finally, an analysis in the microcanonical state has also contributed to the recent debate about the foundations of the canonical formalism \cite{reimann}.

The microcanonical ensemble implies a description of an isolated system. Therefore, one might ask how a relation between fluctuations and dissipation can be possible in a situation where no energy can be dissipated. 
In the present work, our goal is to explore the relation between fluctuations and response in microcanonical equilibrium conditions through the framework of linear response theory. As will be explained later, mainly after the development linear response theory, the name Fluctuation-Dissipation Theorem was associated with some relations which are analogs of the results presented here in the context of the microcanonical ensemble. That is the reason we took the freedom to call them also a FDT even in a situation where there is no physical mechanism for dissipation.
The paper is organized as follows. In Sec. II the derivation of a FDT using linear response theory is presented and its validity is verified in a simple example. In Sec. III different dispersion relations and sum rules are derived in analogy with the usual Kramers-Kronig ones and their meaning is discussed. They are different because they are not derived in the frequency space, like the usual ones. Conclusions are presented finally in Sec. IV.

\section{Derivation of the Fluctuation-Dissipation Theorem}

We start by considering a system whose dynamics is given by a Hamiltonian $\hat{H}$. An external force $K(t)$ is applied to this system such that $\hat{H}$ is now perturbed by an external potential given by $-\hat{A}K(t)$. Following \cite{kubo1}, the reponse function of the system due to the external force measured through an observable $\hat{B}$ is given, in linear response, by
\begin{eqnarray}
 \phi_{BA}(\lambda,t-t')&=&\mathrm{Tr}\left(\hat{\rho}_e(\lambda) 
 \frac{1}{i\hbar}\left[\hat{A}(0),\hat{B}(t-t')\right]\right) \nonumber\\
 &=&\mathrm{Tr}\left(\hat{\rho}_e(\lambda) 
 \frac{1}{i\hbar}\left[\hat{A}(t'),\hat{B}(t)\right]\right),
\label{eq.1}
\end{eqnarray}
where $[\,,\,]$ is the commutator and $\hat{\rho}_e(\lambda)$ is the equilibrium density operator as a function of a macroscopic parameter $\lambda$. One can also define the following correlation function between $\hat{A}$ and $\hat{B}$
\begin{eqnarray}
 C_{BA}(\lambda,t-t')&=&\mathrm{Tr}\left(\hat{\rho}_e(\lambda) 
 \frac{1}{2}\{\hat{A}(0),\hat{B}(t-t')\}\right) \nonumber\\
 &=&\mathrm{Tr}\left(\hat{\rho}_e(\lambda) 
 \frac{1}{2}\{\hat{A}(t'),\hat{B}(t)\}\right),
\label{eq.2}
\end{eqnarray}
where $\{\,,\,\}$ is the anticommutator. This function gives the spectrum of equilibrium fluctuations when the system is unperturbed. For the canonical ensemble, $\hat{\rho}_e(\lambda)=\hat{\rho}_e(\beta)=e^{-\beta\hat{H}}/Z(\beta)$, where $\beta=(k_{B}T)^{-1}$, and the FDT establishes a relation between the spectra of $\phi_{BA}$ and $C_{BA}$. That means a relation between an equilibrium and a nonequilibrium quantity.

Our goal here is to show that there is also a relation between $\phi_{BA}$ and $C_{BA}$ in the microcanonical ensemble. First of all, let us start with the expression for the microcanonical density operator $\hat{\rho}_e(\lambda=E)$. Following \cite{lawson}, we take it as
\begin{eqnarray}
 \hat{\rho}_e(E)=\frac{\delta(E-\hat{H})}{Z(E)},
\label{eq.3}
\end{eqnarray}
where $Z(E)=\mathrm{Tr}\,\delta(E-\hat{H})$.

To derive the FDT, it is necessary to introduce an appropriate representation of $\delta(E-\hat{H})$ like, for example \cite{lawson},
\begin{equation}
 \delta(E-\hat{H})=\frac{1}{2\pi i}\int_{\gamma-i\infty}^{\gamma+i\infty}
  dz\exp{\left[(E-\hat{H})z\right]}.
\label{eq.4}
\end{equation}
Expressions (\ref{eq.1}) and (\ref{eq.2}) can be written now in the following way
\begin{eqnarray}
 \phi_{BA}(E,t-t')=\frac{1}{Z(E)}\mathrm{Tr}\left( \frac{1}{2\pi i}\int_{\gamma-i\infty}^{\gamma+i\infty}
 dz\,e^{(E-\hat{H})z}\,\frac{\left[\hat{A}(t'),\hat{B}(t)\right]}{i\hbar}\right),
\label{eq.5}\\
 C_{BA}(E,t-t')=\frac{1}{Z(E)}\mathrm{Tr}\left( \frac{1}{2\pi i}\int_{\gamma-i\infty}^{\gamma+i\infty}
 dz\,e^{(E-\hat{H})z}\,\frac{\{\hat{A}(t'),\hat{B}(t)\}}{2} \right).
\label{eq.6}
\end{eqnarray}
It is important to note that, since the integrals in the complex plane are always convergent, the trace and integral signs can be interchanged. Doing that, it is convenient to define the following new quantities: $\varphi_{BA}(E,t-t')=Z(E)\phi_{BA}(E,t-t')$ and $\mathcal{C}_{BA}(E,t-t')=Z(E)C_{BA}(E,t-t')$ to obtain
\begin{eqnarray}
 \varphi_{BA}(E,t-t')=\frac{1}{2\pi i}\int_{\gamma-i\infty}^{\gamma+i\infty}
 dz e^{Ez}\chi_{BA}(z,t-t'),
\label{eq.7}\\
 \mathcal{C}_{BA}(E,t-t')=\frac{1}{2\pi i}\int_{\gamma-i\infty}^{\gamma+i\infty}
 dz e^{Ez}F_{BA}(z,t-t'),
\label{eq.8}
\end{eqnarray}
where
\begin{eqnarray}
 \chi_{BA}(z,t-t')=\mathrm{Tr}\left(e^{-\hat{H}z}\frac{\left[\hat{A}(t'),\hat{B}(t)\right]}{i\hbar}\right),
\label{eq.9}\\
 F_{BA}(z,t-t')=\mathrm{Tr}\left(e^{-\hat{H}z}\frac{\{\hat{A}(t'),\hat{B}(t)\}}{2} \right).
\label{eq.10}
\end{eqnarray}
Since $\varphi_{BA}$ and $\mathcal{C}_{BA}$ are given as inverse Laplace transforms of $\chi_{BA}$ and $F_{BA}$, they also satisfy the following relations
\begin{eqnarray}
 \chi_{BA}(z,\tau)=\int_{0}^{\infty}dE\,e^{-Ez} \varphi_{BA}(E,\tau),
\label{eq.11}\\
 F_{BA}(z,\tau)=\int_{0}^{\infty}dE\,e^{-Ez} \mathcal{C}_{BA}(E,\tau),
\label{eq.12}
\end{eqnarray}
where $\tau=t-t'$. We introduce now the Fourier transform of $\chi_{BA}$ and $F_{BA}$,
\begin{eqnarray}
 \tilde{\chi}_{BA}(z,\omega)=\frac{1}{2\pi}\int_{-\infty}^{\infty}d\tau e^{-i\omega\tau}\chi_{BA}(z,\tau),
\label{eq.13}\\
 \tilde{F}_{BA}(z,\omega)=\frac{1}{2\pi}\int_{-\infty}^{\infty}d\tau e^{-i\omega\tau} F_{BA}(z,\tau),
\label{eq.14}
\end{eqnarray}
and also the auxiliary function
\begin{equation}
 S_{AB}(z,\tau)=\mathrm{Tr}\left(e^{-\hat{H}z}\hat{A}(t')\hat{B}(t)\right).
\label{eq.16}
\end{equation}

Noticing that $e^{-\hat{H}z}\hat{A}(t')=\hat{A}(t'+iz\hbar)e^{-\hat{H}z}$ and using the cyclic property of the trace, we obtain
\begin{eqnarray}
 \mathrm{Tr}\left[e^{-\hat{H}z}\hat{B}(t)\hat{A}(t'+iz\hbar)\right]=
\mathrm{Tr}\left[e^{-\hat{H}z}\hat{A}(t')\hat{B}(t)\right].
\label{eq.17}
\end{eqnarray}
Using
\begin{eqnarray}
 \frac{1}{2\pi}\int_{-\infty}^{\infty}d\tau\,e^{-i\omega\tau}\,\mathrm{Tr}
\left[e^{-\hat{H}z}\hat{B}(t)\hat{A}(t'+iz\hbar)
\right]=\frac{1}{2\pi}\int_{-\infty}^{\infty}d\tau'\,e^{-i\omega\tau'}\,
\mathrm{Tr}\left[e^{-\hat{H}z}\hat{B}(t)\hat{A}(t'')\right]e^{z\hbar\omega},
\label{eq.18}
\end{eqnarray}
where $t''=t'+iz\hbar$ and $\tau'=t-t''$, one obtains from (\ref{eq.16}) and (\ref{eq.17})
\begin{eqnarray}
 \tilde{S}_{AB}(z,\omega)=\tilde{S}_{BA}(z,\omega)e^{z\hbar\omega},
\label{eq.19}
\end{eqnarray}
where
\begin{eqnarray}
 \tilde{S}_{BA}(z,\omega)=\frac{1}{2\pi}\int_{-\infty}^{\infty}d\tau'\,e^{-i\omega\tau'}\,
\mathrm{Tr}\left[e^{-\hat{H}z}\hat{B}(t)\hat{A}(t')\right].
\label{eq.20}
\end{eqnarray}
Using (\ref{eq.19}) in the Fourier transforms of (\ref{eq.13}) and (\ref{eq.14}) yields
\begin{eqnarray}
 \tilde{\chi}_{BA}(z,\omega)&=&\frac{1}{i\hbar}\left[\tilde{S}_{AB}(z,\omega)-\tilde{S}_{BA}(z,\omega)\right]=
 \tilde{S}_{BA}(z,\omega)\frac{\left(e^{z\hbar\omega}-1\right)}{i\hbar},
\label{eq.21}\\
 \tilde{F}_{BA}(z,\omega)&=&\frac{1}{2}\left[\tilde{S}_{AB}(z,\omega)+\tilde{S}_{BA}(z,\omega)\right]=
 \tilde{S}_{BA}(z,\omega)\frac{\left(e^{z\hbar\omega}+1\right)}{2}.
\label{eq.22}
\end{eqnarray}
Finally, from (\ref{eq.21}) and (\ref{eq.22}), we obtain
\begin{equation}
 \tilde{F}_{BA}(z,\omega)=i\frac{\hbar}{2}\coth{\left(\frac{z\hbar\omega}{2}\right)}\tilde{\chi}_{BA}(z,\omega),
\label{eq.23}
\end{equation}
which is our quantum FDT. In the classical limit $\hbar\rightarrow 0$, we obtain
\begin{eqnarray}
 \tilde{F}_{BA}(z,\omega)=\frac{i}{z\omega}\tilde{\chi}_{BA}(z,\omega),
\label{eq.24}
\end{eqnarray}
which is our classical FDT. One easily realizes from (\ref{eq.23}) and (\ref{eq.24}) that the replacement of $z$ by $\beta$ in those equations leads precisely to the quantum and classical versions of the FDT in the canonical ensemble. However, the physical nature of (\ref{eq.23}) and (\ref{eq.24}) can be quite different from that in the canonical case. Let us consider, for example, in the classical regime an ergodic and small system, small in the sense that it is not in the thermodynamic limit. Then the microcanonical ensemble averages in (\ref{eq.1}) and (\ref{eq.2}) can be replaced by time averages whose behaviors are given by the dynamics of the system. Therefore, the fluctuations in this case happen due to the dynamics of the concerned system itself and not due to the coupling to a thermostat as in the canonical ensemble. From this point of view, it is surprising that there is a simple relation between the FDT in the canonical and microcanonical ensembles. Indeed, if one wants to compare both cases, the  inverse Laplace transform in $z$ should be performed on (\ref{eq.23}) and (\ref{eq.24}) since the canonical FDT consists of a relation between $\tilde{\phi}_{BA}(\beta,\omega)$ and $\tilde{C}_{BA}(\beta,\omega)$, keeping the original macroscopic parameter $\beta$. For the classical case, this can be easily done using (\ref{eq.24}), leading to
\begin{equation}
 \tilde{\mathcal{C}}_{BA}(E,\omega)=\frac{i}{\omega}\int_{0}^{E}dE'\,\tilde{\varphi}_{BA}(E',\omega).
\label{eq.24b}
\end{equation}
For the quantum regime, the inverse Laplace transform should be performed on (\ref{eq.23}). It is not hard to imagine how different the result will also be from the canonical case.

In addition to the pure meaning of the relation between response and fluctuations, one may wonder whether (\ref{eq.23}) and (\ref{eq.24}) can be useful or not. We would say they can be useful in situations where the microcanonical ensemble can be applied and the thermodynamic limit is not satisfied. However, what we mean by usefulness is the possibility of applying the FDT in a context very different from the ones considered so far to obtain response functions from correlation functions and vice versa. If by useful one meant to go further and  speak about, e.g., transport coefficients, then one would have to discuss more carefully the linear response theory in the microcanonical ensemble, especially because van Kampen's objections \cite{kampen} can be more trickier in this case. The first objection, concerning the validity of the linearization, could still be answered as usual, we believe, by the argument of the stability of the distribution functions \cite{marconi,kubo2}. The second objection, concerning the origin of the decay of correlation functions which lead to finite transport coefficients, cannot be answered as is done sometimes in the context of the canonical ensemble by coupling to an environment \cite{vliet1,vliet2}. The reason is simple: to use the microcanonical ensemble one assumes an isolated system. A possible answer in this case would be the instability of the dynamics \cite{ruelle,gaspard}. However, the question of what ``dissipation'' would mean in the present context of the microcanonical ensemble would remain. This is because, originally, the name Fluctuation-Dissipation Theorem comes from the fact that part of the Fourier transform of the response function is related to the power dissipated by the system when a time-periodic perturbation is applied to it. But for an isolated system there will be no dissipated power. On the other hand, the Fluctuation-Dissipation Theorem, mainly after linear response theory was developed, has been associated with an equation relating the frequency spectra of the response function and of the corresponding symmetric correlation function. In this sense, (\ref{eq.23}) and (\ref{eq.24}) are analogous to Eq. (6.16) of Ref.\cite{kubo1} for the microcanonical ensemble and therefore we took the freedom of calling them Fluctuation-Dissipation Theorems as well. Although beyond the scope of the present work, a general and deep discussion of the subtle points mentioned above as well as of the linear response theory for the microcanonical ensemble would be of great interest and value.

\subsection{Example: the Harmonic Oscillator}

As an example, we would like to check (\ref{eq.23}) and (\ref{eq.24}) for a simple system whose response and correlation functions are known directly. In order to do that, we choose a simple harmonic oscillator. We consider the case $\hat{A}=\hat{B}=\hat{X}$ where $\hat{X}$ is the position operator. To perform first the calculation in the classical regime, we define the classical analogs of (\ref{eq.5}) and (\ref{eq.6}) as
\begin{eqnarray}
 \varphi(E,t-t')=\int\,dx_{o}dp_{o}\,\delta\left(E-H(x_{o},p_{o})\right)\,\{x(t'),x(t)\}_{o},
\label{eq.25}\\
 \mathcal{C}(E,t-t')=\int\,dx_{o}dp_{o}\,\delta\left(E-H(x_{o},p_{o})\right)\,x(t)x(t'),
\label{eq.26}
\end{eqnarray}
where $\{\,,\,\}_{o}$ is the Poisson bracket with respect to the initial conditions $(x_{o},p_{o})$ and $x(t)$ is the solution of the classical equations of motion for the position. The averages above can be easily performed, leading to
\begin{eqnarray}
 \varphi(E,\tau)=\frac{2\pi}{m\omega_{o}^{2}}\sin{\left(\omega_{o}\tau\right)},
\label{eq.27}\\
 \mathcal{C}(E,\tau)=\frac{2\pi E}{m\omega_{o}^{3}}\cos{\left(\omega_{o}\tau\right)}.
\label{eq.28}
\end{eqnarray}
We can now calculate
\begin{eqnarray}
 \tilde{\chi}(z,\omega)=\frac{1}{2\pi}\int_{-\infty}^{\infty}d\tau\,e^{-i\omega\tau}\int_{0}^{\infty}
dE\,e^{-Ez}\,\varphi(E,\tau),
\label{eq.29}\\
\tilde{F}(z,\omega)=\frac{1}{2\pi}\int_{-\infty}^{\infty}d\tau\,e^{-i\omega\tau}\int_{0}^{\infty}
dE\,e^{-Ez}\,\mathcal{C}(E,\tau).
\label{eq.30}
\end{eqnarray}
The results are
\begin{eqnarray}
 \tilde{\chi}(z,\omega)=-i\omega z \frac{2\pi}{m\omega_{o}^{3}}\tilde{g}(\omega)\int_{0}^{\infty}dE\,e^{-Ez}\,E
\label{eq.31}\\
 \tilde{F}(z,\omega)=\frac{2\pi}{m\omega_{o}^{3}}\tilde{g}(\omega)\int_{0}^{\infty}dE\,e^{-Ez}\,E,
\label{eq.32}
\end{eqnarray}
where $\tilde{g}(\omega)=(1/2\pi)\int_{-\infty}^{\infty}d\tau\,e^{-i\omega\tau}\cos{\left(\omega_{o}\tau\right)}$. Therefore,
\begin{eqnarray}
 \tilde{F}(z,\omega)=\frac{i}{z\omega}\tilde{\chi}(z,\omega), \label{eq.33}
\end{eqnarray}
which agrees with (\ref{eq.24}).

Quantum mechanically, we can calculate directly (\ref{eq.9}) and (\ref{eq.10}) for the harmonic oscillator using the energy eigenbasis
\begin{eqnarray}
 \chi(z,\tau)=\sum_{n}e^{-E_{n}z} \frac{\sin{\left(\omega\tau\right)}}{m\omega_{o}},
\label{eq.34}\\
 F(z,\tau)=\sum_{n}e^{-E_{n}z}E_{n}\frac{\cos{\left(\omega\tau\right)}}{m\omega_{o}^{2}}
\label{eq.35}
\end{eqnarray}
where $E_{n}$ are the energy eigenvalues. Therefore, for the Fourier transform $\tilde{\chi}(z,\omega)$ we obtain
\begin{eqnarray}
 \tilde{\chi}(z,\omega)=\sum_{n}e^{-E_{n}z}\frac{i}{2m\omega_{o}}
\left[\delta(\omega_{o}+\omega)-\delta(\omega_{o}-\omega)\right].
\label{eq.36}
\end{eqnarray}
Using (\ref{eq.23}) and (\ref{eq.36}), we obtain an expression for $\tilde{F}(z,\omega)$. Inverting the Fourier transform, we get
\begin{eqnarray}
 F(z,\tau)=\sum_{n}e^{-E_{n}z}\frac{\hbar}{2}\coth{\left(\frac{z\hbar\omega_{o}}{2}\right)}
\frac{\cos{\left(\omega_{o}\tau\right)}}{m\omega_{o}}.
\label{eq.37}
\end{eqnarray}
Since
\begin{eqnarray}
 \sum_{n}e^{-E_{n}z}E_{n}=\frac{\hbar\omega_{o}}{2}\coth{\left(\frac{z\hbar\omega_{o}}{2}\right)}
\frac{1}{2\sinh{\left(z\hbar\omega_{o}/2\right)}},
\label{eq.38}
\end{eqnarray}
Equation (\ref{eq.37}) can be written as
\begin{eqnarray}
 F(z,\tau)=\sum_{n}e^{-E_{n}z}E_{n}\frac{\cos{\left(\omega_{o}\tau\right)}}{m\omega_{o}^{2}}
\label{eq.39}
\end{eqnarray}
which agrees with (\ref{eq.35}). This verification of (\ref{eq.23}) for the quantum harmonic oscillator is the same as in the canonical ensemble case if $z$ is replaced by $\beta$. However, here (\ref{eq.23}) still has to be transformed back to energy.

\section{Dispersion Relations and Sum Rules}

In the canonical ensemble, it is possible to derive relations between the real and imaginary parts of the Fourier transform of the response function \cite{kubo2,kubo3}. Those are the so-called Kramers-Kronig relations and mainly they express a causality property contained in the response function. In the present case, dispersion relations also hold in the $z$-space because $\phi_{BA}$ and $C_{BA}$ are defined for positive values of energy. Equations (\ref{eq.11}) and (\ref{eq.12}) imply that $\chi_{BA}(z,\tau)$ and $F_{BA}(z,\tau)$ are analytic functions in the half plane $\mathrm{Re}(z)\geq\gamma$, where $\gamma$ is positive. Therefore, in this region
\begin{eqnarray}
 \chi_{BA}(z_{o},\tau)=\frac{1}{2\pi i}\oint dz\,\frac{\chi_{BA}(z,\tau)}{z-z_{o}}.
\label{eq.40}
\end{eqnarray}
Since $\lim_{\vert z\vert \to \infty}\vert\chi_{BA}(z,\tau)\vert=0$, we can close the integration contour with a semicircle in the half plane where $\chi_{BA}(z,\tau)$ is analytic and a line along $\mathrm{Re}(z)=\gamma$ and send the radius to infinity to obtain from (\ref{eq.40}) the relation
\begin{eqnarray}
 \chi_{BA}(y_{o},\tau)=\frac{1}{\pi i}\mathrm{P}\int_{-\infty}^{\infty}dy\frac{\chi_{BA}(y,\tau)}{y-y_{o}},
\label{eq.41}
\end{eqnarray}
where the choices $z=\gamma+iy$ and $z_{o}=\gamma+iy_{o}$ were made. The right-hand side denotes the principal value of the integral. Writing $\chi_{BA}$ in terms of its real and imaginary parts, $\chi_{BA}=\chi_{BA}^{\prime}+i\chi_{BA}^{\prime\prime}$, Eq.(\ref{eq.41}) leads to the following dispersion relations
\begin{eqnarray}
 \chi_{BA}^{\prime}(y_{o},\tau)=\frac{1}{\pi}\mathrm{P}\int_{-\infty}^{\infty}dy
\frac{\chi_{BA}^{\prime\prime}(y,\tau)}{y-y_{o}}, \label{eq.42}\\
 \chi_{BA}^{\prime\prime}(y_{o},\tau)=-\frac{1}{\pi}\mathrm{P}\int_{-\infty}^{\infty}dy
\frac{\chi_{BA}^{\prime}(y,\tau)}{y-y_{o}}. \label{eq.43}
\end{eqnarray}
As it is usually done \cite{kubo3}, from the two relations above, it is possible to derive the moment sum rules, which, in this case, are related to the energy dependence instead of the frequency spectrum. The derivation of such sum rules is sketched in the Appendix. The results for the first three moments are shown below, where the subscript $BA$ was dropped for convenience:
\begin{eqnarray}
 \varphi(0,\tau)&=&\frac{1}{\pi}\int_{-\infty}^{\infty}dy\,\chi^{\prime}(y,\tau),
\label{eq.44}\\
 \varphi^{(1)}(0,\tau)&=&-\frac{1}{\pi}\int_{-\infty}^{\infty}dy\,y\,
 \left[\chi^{\prime\prime}(y,\tau)+\frac{\varphi(0,\tau)}{y}\right],
\label{eq.45}\\
 \varphi^{(2)}(0,\tau)&=&-\frac{1}{\pi}\int_{-\infty}^{\infty}dy\,y^{2}\,
 \left[\chi^{\prime}(y,\tau)+\frac{\varphi^{(1)}(0,\tau)}{y}\right],
\label{eq.46}
\end{eqnarray}
where
\begin{eqnarray}
 \varphi^{(n)}(0,\tau)=\left(\frac{\partial^{n}}{\partial E^{n}}\varphi(E,\tau)\right)\bigg\vert_{E=0}.
\label{eq.47}
\end{eqnarray}
The moment sum rules above are related to the asymptotic expansion of $\chi_{BA}$ with respect to $z$ (which means low-energy behavior). For small values of $z$ (i.e. high-energy behavior), one obtains the following sum rules:
\begin{eqnarray}
 \varphi^{(-1)}(0,\tau)&=&-\frac{1}{\pi}\int_{-\infty}^{\infty}dy\frac{\chi^{\prime\prime}(0,\tau)}{y},
\label{eq.48}\\
\varphi^{(-2)}(0,\tau)&=&\frac{1}{\pi}\int_{-\infty}^{\infty}dy\frac{1}{y^{2}}
 \left[\chi^{\prime}(0,\tau)+\varphi^{(-1)}(0,\tau)\right],
\label{eq.49}\\
 \varphi^{(-3)}(0,\tau)&=&\frac{1}{\pi}\int_{-\infty}^{\infty}dy\frac{1}{y^{3}}
 \left[\chi^{\prime\prime}(0,\tau)+y\varphi^{(-2)}(0,\tau)\right],
\label{eq.50}\\
\end{eqnarray}
where
\begin{eqnarray}
 \varphi^{(-n)}(0,\tau)=\int_{0}^{\infty}dE_{1}\int_{E_{1}}^{\infty}dE_{2}\cdots
 \int_{E_{n-1}}^{\infty}dE_{n}\,\varphi(E,\tau).
\label{eq.51}
\end{eqnarray}
The procedure shown in the Appendix can be repeated as long as the derivatives $\varphi^{(n)}$ and the integrals $\varphi^{(-n)}$ exist to derive higher-order moment sum rules.

As for the sum rules in the frequency space, those above can be used to correct phenomenological expressions for $\varphi(E,\tau)$. For example, if one assumes a functional form for the reponse function with some free parameters with respect to the energy dependence, one could determine them by imposing the sum rules for high- or low-energy behavior. The way to do that in the frequency space is shown, for example, in \cite{kubo2,kubo3}. Since the relations above are valid for any value of $\tau$, one could also have dropped the $\tau$ dependence by setting $\tau=0$. Then, it is easier to understand the meaning and the importance of the sum rules: the $z$ spectrum of $\chi$ is given in terms of static quantities like $\varphi^{(n)}(0,\tau=0)$ and $\varphi^{(-n)}(0,\tau=0)$, which could be calculated quantum mechanically in terms of the commutation relations between $\hat{A}$ and $\hat{B}$ (see, for example, \cite{kubo1}).

\section{Conclusions}

Using linear response theory, we presented a derivation of the Fluctuation-Dissipation Theorem in the microcanonical ensemble in both quantum and classical regimes. The theorem is stated as a relation between the Laplace-Fourier transforms of the response and symmetric correlation functions. Although this relation is very similar to the one derived in the canonical ensemble context, it is valid, for example, in a situation where the fluctuations are very different from thermal ones, namely, fluctuations of an isolated system that is not in the thermodynamic limit. Therefore, the Fluctuation-Dissipation Theorem can be considered as a much more general relation and not constrained just to the context of the canonical ensemble. We believe this result can be very useful to calculate correlation functions from response functions (and vice versa) for systems in the microcanonical ensemble when they are not in the thermodynamic limit. In this sense, as mentioned in \cite{lawson,esposito} (see also the references in \cite{esposito}), the present work can be considered as an additional effort to apply statistical physics to small systems. Moment sum rules were also presented for the energy dependence and they could be useful to correct phenomenological expressions for the response functions.

\appendix*
\section{Derivation of the Sum Rules}

In this appendix we give a brief sketch of how to derive the sum rules presented in sec. III. For a careful derivation and deeper discussion about the subject, we refer to \cite{kubo3}. Our starting point is the function $f(z)$ defined by
\begin{eqnarray}
 f(z)=\int_{0}^{\infty}dE\,e^{-Ez}\,\phi(E),
\label{a.1}
\end{eqnarray}
where $z=\gamma+iy$ is complex with its real part positive and $\phi(E)$ is real. Therefore, $f(z)$ is analytic in the half plane $\mathrm{Re}(z)\geq\gamma$ and it satisfies the following dispersion relations:
\begin{eqnarray}
 f^{\prime}(y_{o})=\frac{1}{\pi}\mathrm{P}\int_{-\infty}^{\infty}dy\,\frac{f^{\prime\prime}(y)}{y-y_{o}},
\label{a.2}\\
 f^{\prime\prime}(y_{o})=-\frac{1}{\pi}\mathrm{P}\int_{-\infty}^{\infty}dy\,\frac{f^{\prime}(y)}{y-y_{o}},
\label{a.3}
\end{eqnarray}
where $f^{\prime}(y_{o})$ and $f^{\prime\prime}(y_{o})$ are the real and imaginary parts of $f(y_{o})$, respectively. From (\ref{a.2}), we can write
\begin{eqnarray}
 f^{\prime}(0)=\frac{1}{\pi}\mathrm{P}\int_{-\infty}^{\infty}dy\,\frac{f^{\prime\prime}(y)}{y}
\label{a.4}
\end{eqnarray}
and from (\ref{a.3}) multiplied by $y_{o}$ we obtain
\begin{eqnarray}
 \lim_{y_{o}\to\infty}y_{o}f^{\prime\prime}(y_{o})=\frac{1}{\pi}
 \int_{-\infty}^{\infty}dy\,f^{\prime}(y).
\label{a.5}
\end{eqnarray}
To calculate the left-hand side of (\ref{a.5}), we go back to (\ref{a.1}) and integrate by parts to obtain
\begin{eqnarray}
 f(z)=\frac{\phi(0)}{z}+\frac{1}{z}\int_{0}^{\infty}dE\,e^{-Ez}\,\phi^{(1)}(E),
\label{a.6}
\end{eqnarray}
where 
\begin{eqnarray}
 \phi^{(n)}(0)=\left(\frac{d^{n}}{dE^{n}}\phi(E)\right)\bigg\vert_{E=0}.
\label{a.7}
\end{eqnarray}
Therefore, 
\begin{eqnarray}
 \lim_{\vert z\vert\to\infty} zf(z)=\phi(0), 
\label{a.8}
\end{eqnarray}
and, from (\ref{a.5}) and (\ref{a.8})
\begin{eqnarray}
 \phi(0)=\frac{1}{\pi}\int_{-\infty}^{\infty}dy\,f^{\prime}(y),
\label{a.9}
\end{eqnarray}
which is the first sum rule. To derive the next one, we define a new function
\begin{eqnarray}
 g(z)=\int_{0}^{\infty}dE\,e^{-Ez}\,\phi^{(1)}(E),
\label{a.10}
\end{eqnarray}
which is analytic again for $\mathrm{Re}(z)\geq\gamma$. Therefore, $g(z)$ obeys the same dispersion relations as $f(z)$. Integrating (\ref{a.10}) by parts yields
\begin{eqnarray}
 g(z)=\frac{\phi^{(1)}(0)}{z}+\frac{1}{z}\int_{0}^{\infty}dE\,e^{-Ez}\,\phi^{(2)}(E),
\label{a.11}
\end{eqnarray}
from which we obtain $\lim_{\vert z\vert\to\infty} zg(z)=\phi^{(1)}(0)$. By the same procedure as before,
\begin{eqnarray}
 \phi^{(1)}(0)=\frac{1}{\pi}\int_{-\infty}^{\infty}dy\,g^{\prime}(y).
\label{a.12}
\end{eqnarray}
Since, from (\ref{a.6}) and (\ref{a.10}),
\begin{eqnarray}
 g(z)=\int_{0}^{\infty}dE\,e^{-Ez}\,\phi^{(1)}(E)=zf(z)-\phi(0)
\label{a.13}
\end{eqnarray}
and
\begin{eqnarray}
 g^{\prime}(y)=-yf^{\prime\prime}(y)-\phi(0),
\label{a.14}
\end{eqnarray}
we obtain, from (\ref{a.12}) and (\ref{a.14}), the second sum rule
\begin{eqnarray}
 \phi^{(1)}(0)=-\frac{1}{\pi}\int_{-\infty}^{\infty}dy\,y\left[f^{\prime\prime}(y)+\frac{\phi(0)}{y}\right].
\label{a.15}
\end{eqnarray}
Repeating the same procedure again, we obtain
\begin{eqnarray}
 \phi^{(2)}(0)=-\frac{1}{\pi}\int_{-\infty}^{\infty}dy\,y^{2}
\left[f^{\prime}(y)+\frac{\phi^{(1)}(0)}{y^{2}}\right],
\label{a.16}
\end{eqnarray}
and so on, as long as the $\phi^{(n)}(0)$ exist.

A similar procedure can be applied to generate a different kind of sum rule \cite{kubo3}. Starting again with (\ref{a.1}), we integrate by parts in a different way now:
\begin{eqnarray}
 f(z)=-\phi^{(-1)}(0)+z\int_{0}^{\infty}dE\,e^{-Ez}\,\phi^{(-1)}(E),
\label{a.17}
\end{eqnarray}
where
\begin{eqnarray}
 \phi^{(-n)}(0)=\int_{0}^{\infty}dE_{1}\int_{E_{1}}^{\infty}dE_{2}\cdots
 \int_{E_{n-1}}^{\infty}dE_{n}\,\phi(E).
\label{a.18}
\end{eqnarray}
From (\ref{a.2}),
\begin{eqnarray}
 f^{\prime}(0)=\frac{1}{\pi}\int_{-\infty}^{\infty}dy\,\frac{f^{\prime\prime}(y)}{y},
\label{a.19}
\end{eqnarray}
and from (\ref{a.18}), $f^{\prime}(0)=-\phi^{(-1)}(0)$, so
\begin{eqnarray}
 \phi^{(-1)}(0)=-\frac{1}{\pi}\int_{-\infty}^{\infty}dy\,\frac{f^{\prime\prime}(y)}{y}.
\label{a.20}
\end{eqnarray}
We again repeat the procedure, as before, defining from (\ref{a.20}) a new function $g(z)$,
\begin{eqnarray}
 g(z)&=&\int_{0}^{\infty}dE\,e^{-Ez}\,\phi^{(-1)}(E)=-\phi^{(-2)}(0)
  +z\int_{0}^{\infty}dE\,e^{-Ez}\,\phi^{(-2)}(E)\nonumber\\
  &=&\frac{f(z)}{z}+\frac{\phi^{(-1)}(0)}{z}.
\label{a.21}
\end{eqnarray}
Since $g(z)$ satisfies the same dispersion relations as $f(z)$, we obtain
\begin{eqnarray}
 g^{\prime}(0)=-\phi^{(-2)}(0)=\frac{1}{\pi}\int_{-\infty}^{\infty}dy\,\frac{g^{\prime\prime}(y)}{y}.
\label{a.22}
\end{eqnarray}
Inserting the imaginary part of the second line of (\ref{a.21}) in (\ref{a.22}) leads to
\begin{eqnarray}
 \phi^{(-2)}(0)=\frac{1}{\pi}\int_{-\infty}^{\infty}dy\,\frac{1}{y^{2}}\left[f^{\prime}(y)+\phi^{(-1)}(0)\right].
\label{a.23}
\end{eqnarray}
Repeating the same procedure again, we obtain
\begin{eqnarray}
 \phi^{(-3)}(0)=\frac{1}{\pi}\int_{-\infty}^{\infty}dy\,\frac{1}{y^{3}}
\left[f^{\prime\prime}(y)+y\phi^{(-2)}(0)\right],
\label{a.24}
\end{eqnarray}
and so on, as long as the $\phi^{(-n)}(0)$ exist.

\begin{acknowledgments}
The author acknowledges support of the Brazilian research agency CNPq and DFG (GRK 638). The author is also grateful to J.D. Urbina, M.A.M. de Aguiar and K. Richter for their careful reading of the manuscript and valuable suggestions. Special thanks to R.N.. 
\end{acknowledgments}


\end{document}